\documentclass{article}
\usepackage{latexsym}
\usepackage{amsfonts}
\usepackage{amsmath}
\usepackage{amssymb}
\usepackage{graphicx}
\usepackage{authblk}

\title{\textbf{Axial form factors for quasi elastic $\nu-N$ interactions in the vector/pseudovector dominance model.}}

\author{K.~Kanshin\thanks{k.kanshin@gmail.com} \ and G.~Vereshkov\thanks{gveresh@gmail.com}}

\affil{\emph{The Research Institute of Physics, Southern Federal University, Rostov-on-Don, 344090, Russia}}

\date{}

\begin{document}

\maketitle

\begin{abstract}
Phenomenological multigauge model of neutrino-nucleon interaction based on chiral symmetry of strong interactions and vector/pseudovector meson dominance model is suggested. It was shown  that within the framework of the model the constant of neutron beta decay is formed by meson masses and parameters of the hadronization of fundamental vector bosons. Then, quasi elastic (anti)neutrino-nucleon scattering processes have been investigated and it was found that weak nucleon form factor~(FF) $G_A(Q^2)$ considered as multipole expansion multiplied on pQCD asymptotes is in a good agreement with experimental data.
\end{abstract}

\section{Introduction}
Neutrinos give the unique opportunity to probe axial structure of nucleons, that is why clear theoretical interpretation of experimental data on axial nucleon FFs is of great importance. Moreover, precise knowledge of weak nucleon FFs is vital for present and future accelerator and atmospheric neutrino experiments.

In this work we suggest phenomenological baryon-meson gauge theory of $\beta$-decay and charged current quasi elastic (CCQE) $\nu-N$ scattering. In particular, we introduce multipole expansions of CCQE form factors and compare the results with available experimental data. We studied only free nucleon processes, nuclear effects were not taken into account.

CCQE processes are parameterized by four functions of $Q^2=-q^2$: two vector $F_{1},F_{2}$ and two axial $G_{1},G_{3}$ form factors. However vector FFs are not independent, they can be expressed in terms of ones from elastic electron-nucleon scattering. This relation originates from conserved vector current hypotheses (CVC) \cite{book_gaillard}:
\begin{equation}\label{relFF}
F_a(Q^2)=F^{ep}_a(Q^2)-F^{en}_a(Q^2), \quad a=1,2.
\end{equation}
So far as this work is just a generalization of the theory of $e-N$ elastic scattering presented in \cite{verlal} to the neutrino processes we consider vector weak FFs as well determined.
Besides, contribution of $G_3$ to the cross section is proportional to lepton mass, so in case of light leptons (non $\tau$-lepton processes) it is usually neglected. Thus, in this paper we will focus on only $G_1(Q^2)$.

\section{Basics and features}

The main assumption of the model is the vector and axial vector \textbf{meson dominance}. In case of charged current $\nu-N$ interaction it claims that intermediate vector $W^{\pm}$ bosons hadronize to the sets of vector and axial mesons, so interaction between neutrino and nucleon occurs via meson exchange.

In order to account the sets of mesons with different masses we use \textbf{multigauge approach}. Use of multigauge groups allows to introduce a set of gauge fields with their own coupling parameters, for example in Dirac terms of lagrangian it means:
\begin{displaymath}
\textrm{Dirac:} \qquad
g \bar{\Psi} \gamma^\mu \Psi A_\mu \xrightarrow{multiplifying} g \sum_{i=1}^N w_{i}
\bar{\Psi} \gamma^\mu \Psi A^{i}_\mu,
\end{displaymath}
where
$g$ is a coupling constant,
$\Psi$ is any fermionic field,
$i$ is generation (family) index,
$N$ is a total number of generations (will be discussed later),
$A^{i}_\mu$ are multigauge fields and
$w_i$ - are actual parameters of the theory. Following relation is required to save gauge invariance:
\begin{equation}\label{SR1}
\sum_{i=1}^N w_i =1.
\end{equation}
This is the first sum rule (SR) imposed on parameters.
Moreover, to account anomalous magnetic dipole moment of nucleon Pauli terms are necessary:
\begin{displaymath}
\textrm{Pauli:} \qquad
g \bar{\Psi} \sigma^{\mu \nu} \Psi A_{\mu \nu}
\xrightarrow{multiplifying}
g \sum_{i=1}^N \varkappa_{i} \bar{\Psi} \sigma^{\mu \nu} \Psi A^{i}_{\mu \nu},
\end{displaymath}
however there is no similar relation for  $\sum \varkappa_{i}$ motivated by gauge symmetries. Thus we have two types of parameters: Dirac $w_{i}$ and Pauli $\varkappa_{i}$.

Full theory of both neutral and charged current $\nu- N$ interactions involves at least four sets of mesons which could be classified by $U^{i}_L(1) \times SU^{i}_L(2) \times U^{i}_R(1) \times SU^{i}_R(2)$ isotopic chiral groups. In case of charged currents, only sets of isovector mesons $\rho^{i}$ and $a_1^{i}$ corresponding to
$SU^{i}_L(2)~ \times~ SU^{i}_R(2)$  have to be taking into consideration, while isoscalar  $\omega^{i}$ which are formed by $U^{i}_L(1) \times U^{i}_R(1)$ gauge fields act only in elastic $\nu-N$ and $e-N$ scattering.

Finally, \textbf{chiral symmetry} of strong interactions requires coupling constants of left and right fields to be equal and this way reduces number of parameters. Moreover, it leads to equality of $\rho-N$ and $a_1-N$ coupling. Consequently, parameters of vector and axial form factors will be the same.

By adding nucleon-meson terms to the standard model lepton doublets we receive fermionic sector of lagrangian. In order to construct effective lepton-meson vertexes one has to introduce Higgs fields as a representation of both mesonic and electro-weak gauge groups. To generate meson masses additional Higgs fields are required. Summary of the objects of the theory is given in the table:
\begin{center}
\begin{displaymath}
\begin{array}{|c|c|c|}
  \hline
  \textrm{Field} & \textrm{Group} & \textrm{Vacuum shift} \\
  \hline
N=\dbinom{p}n
&
U^{i}_L(1) \times SU^{i}_L(2) \times U^{i}_R(1) \times SU^{i}_R(2) & \\
l=\dbinom{e} {\nu_e} \ \textrm{or} \ \dbinom{\mu} {\nu_\mu}
&
U^{EW}(1) \times SU_L^{EW}(2)  & \\
  \hline
  \hline
\textrm{Higgs sector}
&
\begin{array}{c}
SU^{EW}_L(2) \times U^{i}_L(1) \times SU^{i}_L(2)                           \\[5pt]
U^{EW}(1) \times U^{i}_R(1) \times SU^{i}_R(2)                               \\[5pt]
U^{i}_L(1) \times SU^{i}_L(2) \times U^{i}_R(1) \times SU^{i}_R(2)       \\[5pt]
U^{i}_L(1)  \times U^{i}_R(1)                                            \\[5pt]
U^{EW}(1) \times SU_L^{EW}(2)
\end{array}  &
\begin{array}{c}
\sim  m_\rho^2                           \\[5pt]
\sim  m_\rho^2                              \\[5pt]
\sim  m_{a_1}^2 - m_\rho^2      \\[5pt]
\sim  m^2_f- m_\omega^2          \\[5pt]
\sim  m_W^2
\end{array} \\
 \hline
\end{array}
\end{displaymath}
\end{center}

Assuming chiral symmetry of the vacuum we put vacuum shifts of left and right (1st and 2nd in the table) Higgs fields equal. Spontaneous symmetry breaking generates quadratic form of boson fields. Its diagonalization leads to the small mixing (only charged bosons considered):
\begin{equation}\label{mixing}
W^{\pm}= W^{\pm} + \varepsilon^{i} \left(\rho^{\pm i}+a_1^{\pm i} \right),
\end{equation}
where $\varepsilon^{i} \sim \dfrac{m_\rho^{i2}}{m_W^2} \ll 1$. This way lepton-meson vertex is introduced.

The remaining degrees of freedom have to be mixed with the Standard Model Higgs boson and then identified with additional meson sets. However lepton-Higgs vertex is suppressed by enormous (245 GeV) value of vacuum shift, so contribution of these mesons is expected to be small and it has been neglected.

After lagrangian had been constructed one could study real physical processes.
\section{$\beta$-decay}
Standard expression for squared $\beta$-decay amplitude which originates from Fermi theory of weak interactions is following:
\begin{equation}\label{M^2 Fermi}
|\mathcal{M}|^2= 16 G^2_F \cos \theta_C (1+3 \alpha).
\end{equation}
Here $G_F=\dfrac{g^2_W} {4\sqrt{2}m_W^2}$ is the Fermi constant, where $g_W$ is the coupling constant of $SU^{EW}_L(2)$, $\theta_C$ is the Cabibbo mixing angle, $\alpha=1.2689$ \cite{pdg} is a dimensionless phenomenological parameter, precisely determined by experiments.

Squared amplitude has been calculated within the framework of the theory as well and parameters of the theory has been identified with ones from the Fermi theory. Thus, $G_F$ is formed by mixing parameters, coupling constants of gauge groups and meson propagators at $Q^2=0$; $\cos \theta_C$  must be inserted due to hadronization via non-perturbative $u\bar{d}$ twist, and finally we have got the following expression for $\alpha$:
\begin{equation}\label{alpha}
\alpha = \sum_{i=1}^N w_{i} \frac{m^2_{\rho_i}}{m^2_{a_{1i}}}
\end{equation}

Taking into account this relation and the fact that parameters $w_i$ are the same for both electron and CC neutrino parts of the theory one had to refit data on electron scattering. This has been done and no any significant changes occurred, parameters have not changed a lot.

\section{$\nu-N$ scattering and the axial FF}
As we have already mentioned the main purpose of studying CCQE scatterng was fitting of $G_1(Q^2)$.
At low $Q^2$ region the conventional dipole form of axial form factor reveals satisfactory fit of the experimental data:
\begin{equation}\label{dipole}
G_1^{dipole}(Q^2)=\frac{\alpha}{ \left(1+\frac{Q^2}{M_A^2} \right)^2 }
\end{equation}
It is parameterized by the  $\beta$-decay constant $\alpha$ and the phenomenological parameter $M_A$, the so-called axial-vector dipole mass.

Within the framework of presented model we have obtained the following multipole expressions of form factors:
\begin{equation}\label{FFs}
\begin{array}{c}
    \begin{array}{ccc}
   \widetilde{F}_1(Q^2)=\sum\limits_{i=1}^N \dfrac{w_i \ m^2_{\rho_{i}} }{m^2_{\rho_{i}}+Q^2} & \qquad  &
   \widetilde{F}_2(Q^2)=\sum\limits_{i=1}^N \dfrac{\varkappa_i \ m^2_{\rho_{i}} }{m^2_{\rho_{i}}+Q^2}
   \end{array}
   \\[10pt]
   \widetilde{G}_1(Q^2)=\sum\limits_{i=1}^N \dfrac{w_i \ m^2_{a_{1i}} }{m^2_{\rho_{i}}+Q^2}           \\
\end{array}
\end{equation}

To reproduce the correct asymptotic behavior of the form factors at infinity the additional sum rules are required:
\begin{equation}\label{SR2}
\begin{array}{rcc}
   F_1(Q^2), \ G_1(Q^2) \sim \dfrac{1}{Q^4}  \qquad & \to \qquad
   &
   \sum\limits_{i=1}^N w_i \ m^2_{\rho_{i}} = 0 \\[20pt]
   F_2(Q^2) \sim \dfrac{1}{Q^6}  \qquad & \to \qquad
   &
   \begin{array}{c}
    \sum\limits_{i=1}^N \varkappa_i \ m^2_{\rho_{i}} = 0\\[10pt]
    \sum\limits_{i=1}^N \varkappa_i \ m^4_{\rho_{i}} = 0
   \end{array}              \\
\end{array}
\end{equation}

Besides, from (\ref{relFF}) and $F^p_2(0)=\mu_p-1, \ F^n_2(0)=\mu_n$, where
$\mu_p$ and $\mu_n$ are magnetic moments of proton and neutron correspondingly we get:
\begin{equation}\label{SR3}
\sum\limits_{i=1}^N \varkappa_{i} = \mu_p - \mu_n -1 \simeq 3.70
\end{equation}

Now let us discuss the number of meson generations $N$. We have obtained six sum rules (\ref{SR1}), (\ref{alpha}), (\ref{SR2}), (\ref{SR3}) which have to be imposed on parameters. To satisfy them at least three generations must be taken into account. The first reason to use only three families is that there is no precise data on heavy axial mesons. Another one is that contribution of $i>3$ mesons is suppressed by increasing mass. And finally, three generations fit of electron data was accurate enough. In the case of $N=3$ both sets of parameters $w_i$ and $\varkappa_i$ $(i=1,2,3)$ are determined from six sum rules. Thus, we consider only three generations, however in the theory there is no any restrictions on number $N$. These mesons are following:
\begin{equation}
\begin{array}{ccc}
\rho(770) & \quad & a_1(1260) \\
\rho(1450) & \quad & a_1(1640) \\
\rho(1700) & \quad & a_1(1930) \\
\end{array}
\end{equation}

To account pQCD asymptotes in high $Q^2$ region one has to multiply form factors (\ref{FFs}) on logarithmic functions $q_a(Q^2)$ of $Q^2$ calculated for Dirac $a=1$ and Pauli $a=2$ FFs in
\cite{Brodsky1974}, \cite{Lepage:1980fj}, \cite{Belitsky:2002kj}, \cite{Brodsky:2003pw}.
The general expression is following:
\begin{equation}
q_a(Q^2)=\left(1+h_a \ln (1+Q^2/\Lambda^2)+g_a \ln^2 \left(1+ Q^2/\Lambda^2 \right) \right)^{-p_a / 2},
\end{equation}
where all the parameters $h_a$ $g_a$ $p_a$ are the same with ones used in electron part of theory. So, they have been already determined from electron data.

Thus we obtain final expressions of FFs:
\begin{equation}\label{FFs_final}
\begin{array}{lcr}
   F_1(Q^2)= q_1(Q^2) \widetilde{F}_1(Q^2)  \qquad
   &
   F_2(Q^2)= q_2(Q^2) \widetilde{F}_2(Q^2) \qquad
   &
   G_1(Q^2)= q_1(Q^2) \widetilde{G}_1(Q^2)   \qquad                \\
\end{array}
\end{equation}

\begin{figure}[!t]
\begin{center}
\includegraphics[width=1.0\textwidth]{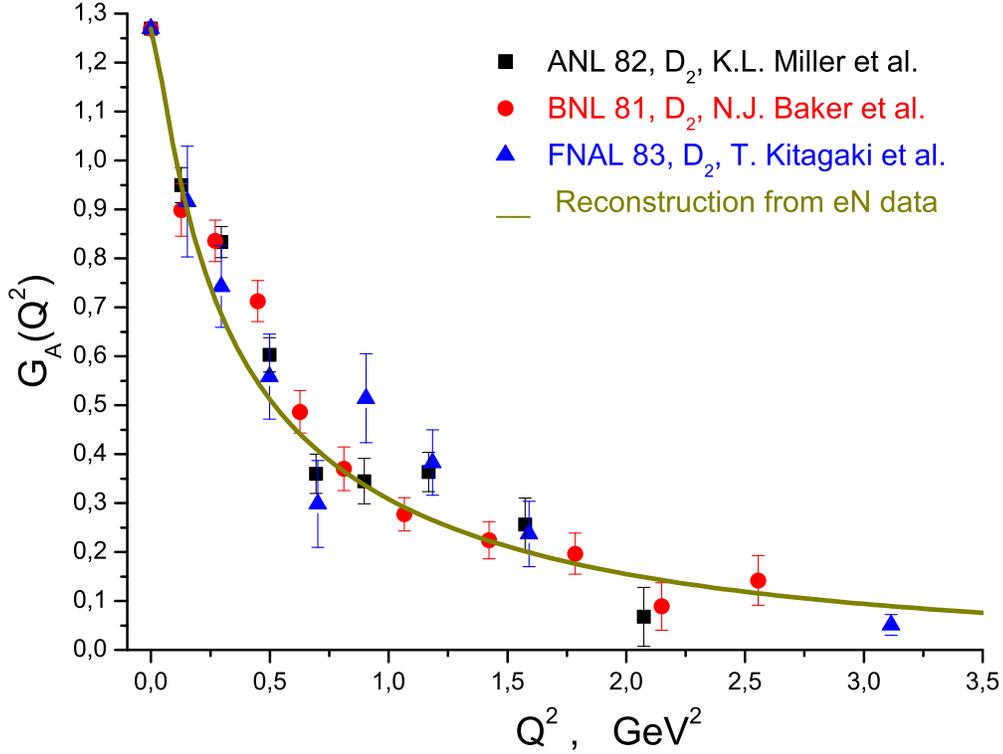}
\caption{Reconstruction of axial formfactor} \label{fig1}
\end{center}
\end{figure}

We emphasize that \emph{there is no free fit parameters} in final FFs. So, we present just a \mbox{\emph{reconstruction}} of $\nu-N$ form factors based on $e-N$ data but not fit. As we have already mentioned nuclear effects was out of consideration. The comparison of $G_1(Q^2)$ from (\ref{FFs_final}) with available deuterium experiments \cite{Miller:1982qi},\cite{Baker:1981su},\cite{Kitagaki:1983px} is presented in the Fig. \ref{fig1}.

\section{Perspectives and conclusion}
We are planning to apply this multigauge scheme based on chiral symmetry and meson dominance to the neutral current as well and thus to complete the full theory of (quasi)elastic lepton-nucleon interactions which consists of three parts:
\begin{enumerate}
  \item Elastic electron-nucleon scattering (presented in \cite{verlal})
  \item Charged current quasi elastic neutrino-nucleon scattering (shortly described here)
  \item Neutral current elastic neutrino-nucleon scattering (under investigation at the moment)
\end{enumerate}

The third part is more complicated then previous ones because of mixing of five boson fields
($\rho^0$,$a_1^0$, $\omega$ - mesons, $Z$-boson and photon). Moreover, in order to introduce effective $\omega$-lepton vertex an additional mixing mechanism is required. It is based on the fact that stress tensor of $U(1)$ groups is gauge invariant, so cross term in the kinematic sector of the lagrangian is not forbidden:
\begin{displaymath}
-\frac{1}{4} \ B_{\mu \nu} \omega^{\mu \nu},
\end{displaymath}
Here $B_{\mu}$ field corresponds to $U^{EW}(1)$ group,
$B_{\mu \nu}= \partial_{\mu} B_\nu - \partial_{\nu} B_\mu$.
Diagonalization of both kinematic and Higgs sectors leads to the required mixing.
The theory of NC will not contain undetermined parameters.

Another perspective application of the approach is a studying of some other processes caused by strong interactions, such as resonance and single pion production or many-pion production.

\end{document}